\documentclass[11pt]{article}

\usepackage{geometry} 
\usepackage{graphicx}
\usepackage{color}
\usepackage{amssymb}
\usepackage{hyperref}
\usepackage{url}
\usepackage{Sweave}
\usepackage{placeins}
\usepackage{caption}
\usepackage{subcaption}
\usepackage{mathtools}
\usepackage[T1]{fontenc}
\usepackage[utf8]{inputenc}
\usepackage{authblk}

\newcommand{\be}{\begin{equation}}
\newcommand{\ee}{\end{equation}}
\newcommand{\bse}{\begin{subequations}}
\newcommand{\ese}{\end{subequations}}
\newcommand{\bea}{\begin{eqnarray}}
\newcommand{\eea}{\end{eqnarray}}

\newcommand{\bs}{\boldsymbol}

\usepackage[default]{jasa_harvard}
\usepackage{JASA_manu}

\newcommand{\itc}[1]{\textcolor{blue}{\textit{#1}}}

\date{}

\title{\textbf{rCOSA: A Software Package for Clustering Objects on Subsets of Attributes}}

\author[1]{Maarten M. Kampert\thanks{
	correspondence: mkampert@math.leidenuniv.nl \\
	Authors' Addresses: \\
	M.M. Kampert, Mathematical Institute, Leiden University, Niels Bohrweg 1, 23333 CA Leiden, \\
	e-mail: mkampert@math.leidenuniv.nl;\\
	J.J. Meulman, Mathematical Institute, Leiden University, Niels Bohrweg 1, 2333 CA Leiden, 
	\\ e-mail: jmeulman@math.leidenuniv.nl;\\
	J.H. Friedman, Department of Statistics, Stanford University, 390 Sarah Mall, Stanford CA 94305, \\e-mail: jhf@stanford.edu.\\	
	}
}

\author[1,2]{Jacqueline J. Meulman}
\author[2]{Jerome H. Friedman}
\affil[1]{Mathematical Institute, Leiden University}
\affil[2]{Department of Statistics, Stanford University}

\date{}

\begin{document}

\maketitle

\setcounter{page}{1}


\newpage
\thispagestyle{empty}

\title{\textbf{rCOSA: A Software Package for Clustering Objects on Subsets of Attributes}}

\abstract{ \texttt{rCOSA} is a software package interfaced to the R language. It implements statistical techniques for clustering objects on subsets of attributes in multivariate data. The main output of COSA is a dissimilarity matrix that one can subsequently analyze with a variety of proximity analysis methods. Our package extends the original COSA software (Friedman and Meulman, 2004) by adding functions for hierarchical clustering methods, least squares multidimensional scaling, partitional clustering, and data visualization. In the many publications that cite the COSA paper by Friedman and Meulman (2004), the COSA program is actually used only a small number of times. This can be attributed to the fact that thse original implementation is not very easy to install and use. Moreover, the available software is out-of-date. Here, we introduce an up-to-date software package and a clear guidance for this advanced technique. The software package and related links are available for free at: \url{https://github.com/mkampert/rCOSA}}



\newpage
\section{Introduction}
Visual representations of dissimilarities (proximities, distances) are advantageous for discovery, identification and recognition of structure in many fields that apply statistical methods. Clustering objects in multivariate (attribute-value) data is a highly popular data analysis objective. Distance-based methods define a measure of similarity, e.g. a composite based on distance derived from each attribute separately.
Let an object $i$ be defined as $o_{i}=\mathbf{x}_{i}=(x_{i1},x_{i2},\ldots,x_{ik})$, 
where $\{x_{ik}\}_{k = 1}^P$ denotes the measured attributes on each object $i$.
\(\mathbf{X}\) denotes the data matrix of size \(N \times P\), with $N$ objects and $P$ attributes (or variables). 
For each attribute $k$, we calculate the distance $d_{ijk}$ between a pair of objects $i$ and $j$ as follows: 
\bea
d_{ijk} = | x_{ik} - x_{jk} | / s_k,
\eea
with $s_k$ a scale factor, a measure of dispersion. If $s_k$ is set as $ \frac{\sigma}{\sqrt{n}}$, with $\sigma$ the standard deviation of $\mathbf{x}_{k}$, then $d_{ijk}$ is the distance between object $i$ and $j$ in the standardized variable $\mathbf{x}_k$.
For categorical attributes, we calculate the \(d_{ijk}\) of object pair \(i\) and \(j\) as  
\bea
d_{ijk} = I( x_{ik} \neq x_{jk}) / s_k,
\eea 
with $s_k$ a suitable scale factor for categorical variables. When all attributes are numeric, and we set \(s_k\) equal to $ \frac{\sigma}{\sqrt{n}}$, 
then the sum of all attribute distances for objects $i$ and $j$ defines the \(L_1\) distance 
\bea
\label{DL1}
D_{ij} = \sum^P_{k = 1} d_{ijk}
\eea
for standardized variables. The squared Euclidean distance would be obtained by taking
\bea
\label{DL2}
D_{ij}^2 = \sum^P_{k = 1} d_{ijk}^2.
\eea

\section{Clustering on subsets of attributes}

\begin{figure}[ht]
\begin{center}
\includegraphics[]{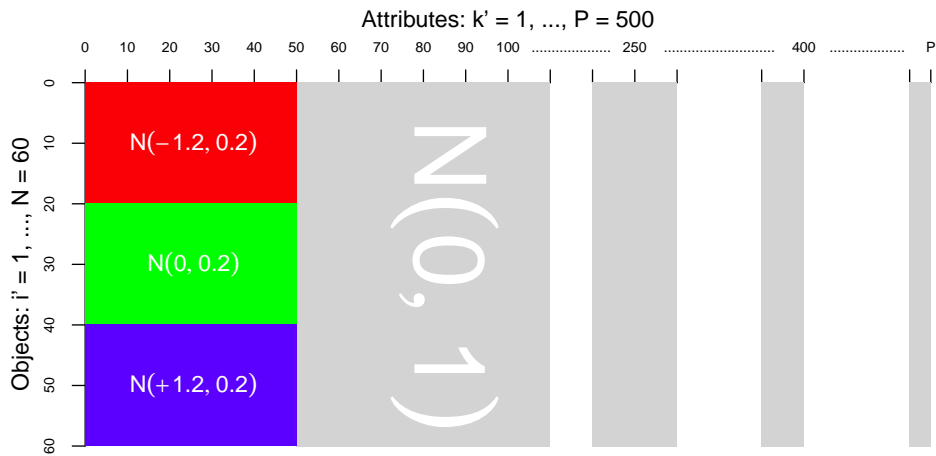}
\caption { A Monte Carlo data set \(\mathbf{X}\) with 60 objects (vertical) and 500 attributes (horizontal, not all of them are shown due to \(P >> N\)). There are three groups of 20-objects each (red, green, and blue) clustering on 50 attributes. Note that \(i\) and \(k\) are ordered into \(i'\) and \(k'\), respectively, to show cluster blocks.}
\label{fig:ExDesign1Intro}
\end{center}
\end{figure}

The focus on clustering of objects on subsets of attributes was motivated by the presence of high-dimensional data, emerging from fields like genomics (e.g., gene expression micro-array data), and metabolomics (e.g., LC-MS data), where the data consist of a very large number of attributes/variables compared to a relatively small number of objects.  Ordinary clustering techniques, based on \eqref{DL1} or \eqref{DL2} use equal weights for each attribute, and this might cause masking of existing clustering, because with a large number of attributes, it is very unlikely that objects cluster on \emph{all} attributes. Instead, objects might be preferentially close on some attributes and far apart on others. This situation calls for feature selection, or assigning a different weight to each attribute; applying a clustering procedure to Euclidean and \(L_1\)-distances does not perform well in general when only a few attributes contain signals and most others contain noise. In such situations, clustering applied to dissimilarities that incorporate variable weighting are much more likely to succeed in finding groups in the data. Figure \ref{fig:ExDesign1Intro} shows a display for a toy-example data set for which it will be unlikely that clustering of either Euclidean or \(L_1\) distances would capture the signal.

\FloatBarrier

For data as displayed in Figure \ref{fig:ExDesign1Intro}, we can expect the clustering procedure to be successful when the dissimilarity measure would incorporate variable selection/weighting. In partitional clustering the weighting of attributes has received considerable attention (For example, De Sarbo, Carroll, Clarck and Green 1984; Steinley and Brusco 2008; Jain 2010; Andrews and McNicholas 2014), but not so for dissimilarity and distance functions. There are studies where attribute weighting is applied, but either these methods are not capable to capture signal in high-dimensional data settings where $P >> N$,  or have as sole purpose to fit a tree in hierarchical clustering (Sebestyen 1962, De Soete, De Sarbo and Carroll 1985; De Soete 1985; Amorim 2015). Sparse clustering (SPARCL) by Witten and Tibshirani (2010) can output an attribute weighted dissimilarity measure for the objects. 

Denote the element \(w_k\) as the weight for attribute dissimilarity $d_{ijk}$, then the composite dissimilarity measure that incorporates variable weights is given in (\ref{SPARCLd})
\bea
\label{SPARCLd}
D_{ij}[\mathbf{w}] &=& \sum^P_{k=1} w_{k}d_{ijk}.
\eea 
As we shall see below, restrictions are needed on the $\mathbf{w}= \{w_k\}$ to prevent degenerate solutions (also, see Witten and Tibshirani (2010)).
%
We will start our discussion with the case were only \emph{one} subset of attributes is important for \emph{all} groups of objects, and where the groups only differ in their means. This particular case was displayed in Figure \ref{fig:ExDesign1Intro}.

\begin{figure}[ht]
\begin{center}
\includegraphics[scale = 0.25]{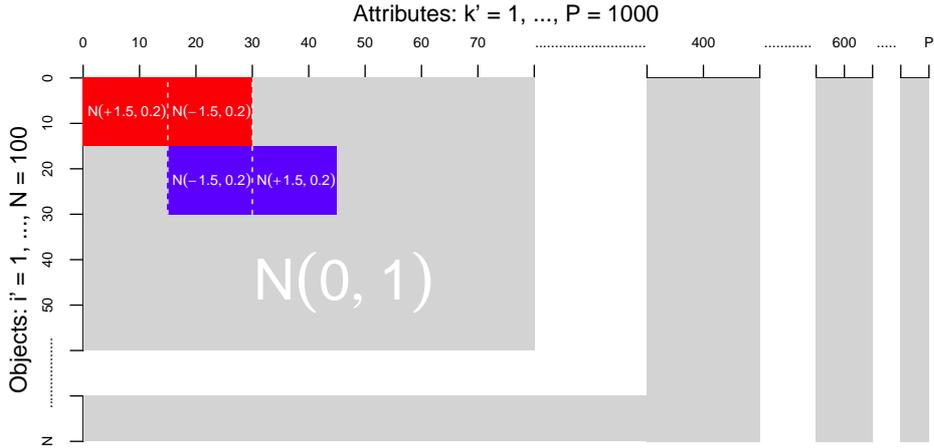} \caption{A Monte Carlo model for 100 objects with 1,000 attributes (not all are shown due to \(P >> N\)). There are two small 15-object groups (red \& blue), clustering each on 30 attributes out of 1000 attributes, with partial overlap, and nested within an unclustered background of 70 objects (gray).\itc{}}
\label{fig:ExDesign2Intro}
\end{center}
\end{figure}

It is important to realize that in this example, \emph{all} objects are assumed to be in clusters; there are no objects in the data that do \emph{not} belong to one of the clusters. 
This is a very particular structure, and is unlikely to be present in many high-dimensional settings. 
In many data sets, one can hope to find one or more clusters of objects, while the remainder of the objects are not close to any of the other objects. Moreover, it could very well be true that one cluster of objects is present in one subset of attributes, while another cluster is present in another subset of attributes. 
In this case, the subsets of attributes are different for each cluster of objects. 
In general, the subsets may be overlapping or partially overlapping, but they may also be disjoint.
An example is shown in Figure \ref{fig:ExDesign2Intro}; the display shows a typical structure in which the groups of objects cluster on their own subset of attributes. The first group (with objects 1-15) clusters on the attributes 1-30, and the second group (with objects 16-30) clusters on attributes 16-45. So the two groups are similar with respect to attributes 16-30, and different with respect to attributes 1-15 and 31-45, respectively. The two subsets of attributes, 1-30 and 16-45, are partially overlapping. The remaining 70 objects in the data form an unclusterable background (noise), and the remaining 955 attributes do not contain any clusters at all.  
The data structure displayed in Figure \ref{fig:ExDesign2Intro} is a typical example for which COSA was designed. 
In such a situation, heuristic, greedy cluster algorithms are very likely to convergence to suboptimal solutions. To avoid such solutions as much as possible, we would need a clever search strategy, together with a good starting point. The latter is crucial: when we start the search with equal weighting of attributes in combination with a usual definition of `closeness' -such as Euclidean or $L_1$ distance- our search will almost surely end up in a distinctly suboptimal local minimum. 

\section{COSA Dissimilarities}
The approach that is used in COSA, is to modify the original cluster criterion, defined on $L_1$ distances \eqref{DL1}, by using a very particular distance instead, for which the equal weights starting point is not detrimental. During the search, in which the optimal weights are found, this particular distance transitions into an ordinary weighted $L_1$ distance. A penalty is used to avoid obtaining subsets that are trivially small (e.g., consisting of a single attribute).  
In this section, we will briefly give the technical details. 



%
%
%
%
Friedman and Meulman (2004) propose an algorithm that uses the \emph{weighted inverse exponential distance}, defined as
\bea
D_{ij}^{(\lambda)}[\mathbf{w}]=-\lambda\log\sum_{k=1}^{P}%
w_{k}\,e^{-d_{ijk}/\lambda},
\eea
where $\lambda$ is a scale parameter, defining ``closeness" between objects.  Because the distance
\bea
D_{ij}^{(\lambda)}[\mathbf{w}] \simeq-\lambda\log\sum_{d_{ijk\lesssim\lambda}}%
w_{k}\,e^{-d_{ijk}/\lambda},
\eea
basically gives emphasis to all distances smaller than a particular value for $\lambda$ for any value of the weights including $w_k = 1/P$. If we define a parameter $\eta$, and define 
\bea
D_{ij}^{(\eta)}[\mathbf{w}]=-\eta\log\sum_{k=1}^{P}w_{k}\,e^{-d_{ijk}/\eta},
\eea
then when $\eta$ increases\ from $\lambda$\ to $\infty$, we obtain a transition from the
weighted inverse exponential distance to the weighted $L_1$ distance, because
\bea
\lim_{\eta\rightarrow\infty}D_{ij}^{(\eta)}[\mathbf{w}]=\sum_{k=1}
^{P}w_{k}d_{ijk}=D_{ij}[\mathbf{w}].
\eea
By using this so-called homotopy strategy, COSA attemps to avoid local minima by starting the iterative process with inverse exponential distances (where equal weighting is not detrimental) that will change into ordinary $L_1$ distances during the process.
 
The weight for a pair of objects is then defined as:
\bea
\label{COSAd}
D_{ij}[\mathbf{W}] &=& \sum^P_{k=1} \max(w_{ik},w_{jk}) d_{ijk}, \nonumber \\
&& \mbox{subject to   } 0 \leq w_{ik} \leq 1 \mbox{ and } \sum^P_{k=1} w_{ik} = 1 \,\, \forall i.
\eea

Object pairs that belong to the same cluster will obtain weights that are more similar compared to object pairs that don't belong to the same cluster. The COSA dissimilarity in \refeq{COSAd} can uncover groups that cluster on their own set of attributes. The larger the difference between \(w_{ik}\) and \(w_{jk}\), the larger the dissimilarity \(D_{ij}[\mathbf{W}]\). 

The COSA weights and the associated dissimilarities are found by minimizing the criterion
\bea
\label{COSAcrit}
Q(\mathbf{W}) = \sum^{N}_{i = 1}\left\{ K^{-1} \sum_{j \in KNN(i)} D_{ij}[\mathbf{w}_i ] + \lambda \sum^P_{k = 1} w_{ik}\log(w_{ik}) \right\}.
\eea
Here, \(K\) is a pre-set number of nearest neighbors, by default set to \(K = floor(\sqrt{N})\). The \(j \in \mbox{KNN}{(i)}\) denotes the \(j = 1 \ldots K\) nearest neighbor objects for object \(i\). The \(\mathbf{w}_{i}\) vector, is the \(i^{th}\) row of \(\mathbf{W}\), and makes the minimization problem linear since the term \(\max(w_{ik},w_{jk})\) is now absent in \eqref{COSAcrit}. Equation \eqref{COSAcrit} is written as a Lagrangian form, an equivalent way to write the equation is 
\bea
\label{COSAcritLAGRANGE}
Q(\mathbf{W}) = \sum^{N}_{i = 1} K^{-1} \sum_{j \in KNN(i)} D_{ij}[\mathbf{w}_i], \nonumber \\ 
\mbox{subject to }\sum^P_{k = 1} w_{ik}log(w_{ik}) \geq t_i \,\,\forall\, i. 
\eea

The Lagrangian term \(t_i\), also called the penalty regularized by \(\lambda\), ensures that subsets of attributes will not be trivially small. The larger  \(\lambda\), the smaller the penalty \(t\), and the more equal the weights for the attributes. Vice versa, the smaller \(\lambda\), the larger the penalty \(t_i\), and hence, the stronger a subset of attributes is favored over others.  

%
%
For known \(j \in \mbox{KNN}{(i)}\) and \(\lambda\), there is an analytical solution for \(\mathbf{W}\) that minimizes \(Q(\mathbf{W})\) in \eqref{COSAcrit} and \eqref{COSAcritLAGRANGE}; to be specific an element $w_{ik}$ is obtained as

\bea
\label{COSAweights}
w_{ik} = \exp \left( - \frac{ \sum_{j \in \mbox{KNN}(i)} d_{ijk} }{K\lambda} \right) \left / \sum^P_{k'=1} \exp \left( - \frac{ \sum_{j \in \mbox{KNN}(i)} d_{ijk'} }{K\lambda} \right) \right . .
\eea
Since the \(D^{(\eta)}_{ij} [\mathbf{W}] \), on which we base the \(j \in \mbox{KNN}{(i)}\), are not known beforehand, we have to  iteratively minimize the criterion. Summarizing, COSA uses a homotopy strategy by starting with the inverse exponential distance 
\bea
\label{invexpd}
D^{(\eta)}_{ij} [\mathbf{W}] = -\eta \log\left\{ \sum^P_{k=1} \max(w_{ik},w_{jk})\exp\left(-\frac{d_{ijk}}{\eta}\right) \right\},
\eea
with \(\eta\) the homotopy parameter.  During the iteration process, the inverse exponential distance transitions into the $L_1$ distance by slowly increasing the value of $\eta$. As is mentioned in Friedman and Meulman (2004), the correlation between the two set of distances is already .91 for $\eta$ =1, and .97 for $\eta$ =2 for distances derived from normally distributed attribute values $w_{ik}$ and equal weights $w_{k}=1/P$. 

Having defined the necessary ingredients, we can now summarize the COSA algorithm in the following six steps:

\FloatBarrier
\begin{center}
\vskip .5cm
\textbf{COSA Algorithm}\\
$%
\begin{array}{llr}
\\
1 & \text{Initialize: } \eta = \lambda; \mathbf{W}=\{1/P\} \in \mathbb{R}^{N \times P}\\
2 & \text{Outer Loop \{} \\ 
3 & \qquad\text{Inner Loop \{} \\ 
 & \qquad\qquad\text{Compute distances }D_{ij}^{\eta}[\mathbf{W}] & \qquad\text{\eqref{invexpd}}  \\
 & \qquad\qquad\text{Compute weights } \mathbf{W} &  \qquad\text{\eqref{COSAweights}}\\
 & \qquad \}\text{ Until convergence.} \\
 4 & \qquad \text{Increase }\eta : \eta + 0.1*\lambda \\
5 & \} \text{ Until }\mathbf{W}\text{ stabilizes} \\ 
6 & \text{Output: }\{D_{ij}^{\eta}[\mathbf{W}], \mbox{ and } \mathbf{W}\}%
\end{array}
$
\end{center}
\FloatBarrier
We refer to Friedman and Meulman (2004) for more details and properties of the algorithm. 

\section{Targeting}
Until now the COSA clustering could be on any possible joint values on subsets of attributes.
Alternatively, we could wish to look for clusters that group only on particular values, 
say $t_{k}$, which are possibly different for each attribute $k$.
The \{$t_{k}$\} are chosen to be of special interest; we reduce the search space, and would hope
to be more likely to recover clusters. Examples are groups of consumers (objects) that spend relatively large amounts on products (attributes), while we wish to ignore consumers who spend relatively small or average amounts. (Or the other way around.) If we focus on one particular value, we call this single targeting. We modify the original distance between objects $o_i$ and $o_j$ on attribute $k$, $d_{ijk}=d_{k}(x_{ik},x_{jk})$, into targeted distances,
and require objects $o_i$ and $o_j$ to be close to each other $\emph{and}$ to the particular target. The so-called single target distance is defined as:
\begin{equation}
d_{ijk}(t_{k})=\max[d_{k}(x_{ik},t_{k}),d_{k}(x_{jk},t_{k})], \label{e15}%
\end{equation}
where $t_{k}$ is the target value, e.g., a $\emph{high}$ or $\emph{low}$ or even $\emph{average}$ value. This
distance is small only if both objects $o_{i}$ and $o_{j}$ are close to the target value $t_{k}$ on attribute $k$. In addition to single targeting, we can also focus on two different targets,
being naturally either high or low values. An example is in micro array data, where we could search for clusters of samples with either high or low (but not moderate) expression levels
on subsets of genes (attributes). In dual targeting, we define two targets $t_{k}$ and $u_{k}$, and we use the dual target distance
\begin{equation} 
d_{ijk}(t_{k},u_{k})=\min[d_{ijk}(t_{k}),d_{ijk}(u_{k})]
\end{equation}
on selected attributes $x_{k}$, where $d_{ijk}(\cdot)$ is the corresponding
single target distance (\ref{e15}). This dual target distance is small
whenever $x_{ik}$ and $x_{jk}$ are either both close to $t_{k}$ \emph{or}
both close to $u_{k}$. Thus, in gene expression and consumer spending examples,
one might set $t_{k}$ and $u_{k}$ to values near the maximum and
minimum data values of the attributes, respectively, and we will cause COSA 
to seek clusters based on extreme attribute values,
ignoring (perhaps dominant) clusters with moderate attribute values.

\section{Installing and using COSA}
In the sequel of this paper, we will present the new version of COSA implemented as a package for the statistical computing language \texttt{R}  (R Core Team 2014). Compared to the old software, the current installation is much simpler, and it is extended with functions for multidimensional scaling, graphics, and \(M\)-groups clustering methods. For every function in \texttt{rCOSA} there is a help file with example code that users can run. The software is available for up-to-date Windows, Mac OSx and Linux platforms.  The \texttt{rCOSA} package, the user manual (vignette) and related links are available for free on the web at: \url{https://github.com/mkampert/rCOSA}. To install rCOSA, run the following code in \texttt{R}.

\vskip 0.5cm
\begin{Sinput}
> install.packages("devtools")
> library(devtools) # for loading the function install_github
> install_github("mkampert/rCOSA") # install rCOSA
> library(rCOSA) # load rCOSA
\end{Sinput}
\section{Using COSA}

We illustrate the rCOSA package using a data set based on the simulation model shown in Figure \ref{fig:ExDesign2Intro}. 
The data set $\mathbf{X}_{N \times P}$ (with $N = 100$ and $P = 1000$) contains two groups, and background noise. The two groups share a subset of 15 attributes, of which each $x_{ik}\sim N( \mu = -1.5, \sigma =  0.2)$. Each groups also has its own unique subset of 15 attributes. These non-overlapping subsets, are $x_{ik} \sim N(\mu = +1.5, \sigma =  0.2)$ for both groups. All the remaining data in $\mathbf{X}$ were generated from a standard normal distribution. After creating the groups, the pooled sample was standardized to have zero mean and unit variance on all attributes. The data set thus contains two small groups that exhibits clustering on only a few non-overlapping and perfectly overlapping attributes, together with a large non-clustered background. 

Possible \texttt{R} code for simulating such a data set, and reproducing our tutorial results is as follows:
\begin{Sinput}
> set.seed(123); N <- 100; P <- 1000;
> X <- matrix(rnorm(N*P), nrow = N, ncol = P)
> i <- sample(x = 1:N) #  i conform Figure 2
> k <- sample(x = 1:P) #  k conform Figure 2
> X[i[1:15], k[1:15]] <- X[i[1:15], k[1:15]]*0.2 + 1.5
> X[i[1:15], k[16:30]] <- X[i[1:15], k[16:30]]*0.2 - 1.5
> X[i[16:30], k[16:30]] <- X[i[16:30], k[16:30]]*0.2 - 1.5
> X[i[16:30], k[31:45]] <- X[i[16:30], k[31:45]]*0.2 +  1.5
> X <- data.frame(scale(X))
\end{Sinput}
\vskip 0.5cm

We run COSA using its default settings and store it in the object \texttt{cosa.rslt} in the following way:
\begin{Sinput}
> cosa.rslt <- cosa2(X) 
\end{Sinput}

In Linux and Mac OS X based operating systems this will start the console to display the output of the successive COSA iterations; for Windows platforms a command prompt window opens in which the following output is shown:
\vskip 0.5cm
\begin{Soutput}
 COSA executing (enter ESC or ctrl+c to terminate).

   Wchange       #iit #oit #it   Eta          MSD              Crit
   0.565417        1    1    1   0.2200       0.789181         87.6947
   0.165954        2    1    2   0.2200       0.803855         85.5847    
   ....
   0.000000        1  100  130   2.200        0.307085E-01     84.9413
\end{Soutput}
The first column indicates the changes in the weights \(\Delta\mathbf{W}\) after each iteration (\#it) defined as the sum of the absolute differences between the weights in \(\mathbf{W}^{(itrs)}\) and the weights in the previous iteration \(\mathbf{W}^{(itrs  -1)}\). The \#iit column gives the number of inner iterations and the \#oit column the number of outer iterations. The eta column shows the value of the homotopy parameter \(\eta\), which starts low and is defined as \(\eta = \lambda + \# oit*0.1*\lambda\). Gradually increasing the homotopy parameter tries to avoid local minima for the criterion (which is of course not guaranteed). The Mean of the Squared Differences (MSD) is defined as

\be
MSD = \frac{ \sum_{j = 1}^{N-1} \sum_{i = j + 1}^N \left( D_{ij}[\mathbf{W}]  - D^{\eta}_{ij}[\mathbf{W}] \right)^2}{ N \cdot ( N - 1)}, 
\ee

and gives the a verage of the squared differences between the $L_1$ distances and the inverse exponential distances. The last column gives the value of the criterion as displayed in equation \eqref{COSAcrit}. The \texttt{R} function \texttt{str()} shows the contents of the output object \texttt{cosa.rslt}.
\vskip 0.5cm

\begin{Sinput}
> str(cosa.rslt)
\end{Sinput}
\begin{Soutput}
List of 4
 $ call  : language cosa2(X)
 $ D     :Class 'dist'  atomic [1:4950] 0.147 0.146 0.125 0.155 0.16 ...
  .. ..- attr(*, "Size")= int 100
  .. ..- attr(*, "Diag")= logi FALSE
  .. ..- attr(*, "Upper")= logi FALSE
 $ W     : num [1:100, 1:1000] 0.21568 0.08017 0.00277 1.18862 1.72863 ...
 $ tunpar:List of 7
  ..$ crit    : num 84.9
  ..$ lambda  : num 0.2
  ..$ homotopy: num 2.2
  ..$ MSD    : num 0.0307
  ..$ Knn     : num 10
  ..$ noit    : num 100
  ..$ totit   : num 130
\end{Soutput}
Thus, the function \texttt{cosa2} gives a list of 4 objects. The first object, \texttt{..\$ call} is an echo of the used \texttt{cosa2} command. The second and third objects in \texttt{cosa.rslt}, are the dissimilarities \texttt{cosa.rslt\$D} and the weights \texttt{cosa.rslt\$W}, respectively. Last, the \texttt{..\$ tunpar} object gives the criterion, $\lambda$ parameter, $\eta$ parameter, MSD, $K$, number of outer iterations, and number of inner iterations.

\subsection{Fitting dendrograms to COSA dissimilarities}

\begin{figure}[ht]
\begin{center}
\includegraphics[width=7cm,keepaspectratio]{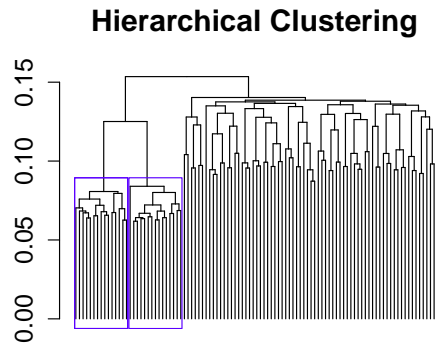}
\caption{Selecting two groups out of the dendrogram of the COSA Dissimilarities using the function \texttt{getclust}}\label{fig:SelGrDendr}
\end{center}
\end{figure}

To display the possible clustering structure contained in the COSA dissimilarities (\texttt{cosa.rslt\$D}), we can first plot a dendrogram using the \texttt{hierclust} function. By default, the dendrogram is build using average linkage. Other options such as `single', `complete', and `ward' linkage are available; the command for ward clustering would be \texttt{hierclust(cosa.rslt\$D, method = `ward')}  (Ward Jr 1963). To ensure that this dendrogram has a scale that is comparable with future dendrograms, the COSA dissimilarities are by default normalized to have sum of squares equal to $N$. 
To plot a dendrogram, use 
\begin{Sinput}
> hclst.cosa <- hierclust(cosa.rslt$D)
\end{Sinput}
From the dendrogam given by the \texttt{hierclust} command, we can clearly see the grouping structure conform to the design that was used. There are two groups (each with 15 objects) and a large remaining group for which the objects are not similar to each other. We can select the observed clusters, and obtain the index numbers of the objects in each cluster, by using 
the \texttt{getclust} function:
\begin{Sinput}
> grps.cosa <- getclust(hclst.cosa)
\end{Sinput}
This function reads the position of the pointer, and with a click we can cut the tree at the vertical position of the pointer, and draw a colored rectangle around the cluster. The index numbers of the objects in the corresponding groups are then stored in the object \texttt{grps.cosa}. When finished, press `Esc' or choose \texttt{Stop} from the options using the right-click of the mouse. Figure \ref{fig:SelGrDendr} shows the two groups we selected. 
The content of \texttt{grps.cosa} can be seen by using the command \texttt{str(grps.cosa)}:
\begin{Soutput}
> str(grps.cosa)
List of 2
 $ grps : num [1:100] 2 0 0 0 0 0 2 2 0 0 ...
 $ index:List of 2
  ..$ grp1: int [1:15] 21 22 23 38 46 47 52 59 61 62 ...
  ..$ grp2: int [1:15] 1 7 8 11 16 19 29 41 55 66 ...
\end{Soutput}
The first line indicates whether an object is from a particular group, and if so, which group label is attached. If an object has not been allocated to a group, it gets a 0. The subsequent lines give the indices for the objects in the selected groups.  

\subsection{Fitting multidimensional scaling solutions to COSA dissimilarities}
In addition to hierarchical clustering producing a dendrogram, we can also use the COSA dissimilarity matrix to display the objects in low-dimensional space by multidimensional scaling (MDS). This is done preferably by using an algorithm that minimizes a least squares loss function, usually called STRESS, defined on dissimilarities and distances. This loss function (in its raw, squared, form) is written as:
\bea
\label{SMACOF}
\mbox{STRESS}(\mathbf{Z})= \left| \left| \bs{\Delta} - \mathbf{D}(\mathbf{Z}) \right| \right|^2,
\eea 
\noindent  where $\left| \left| \cdot \right| \right|^2$ denotes the squared Euclidean norm. Here \(\bs{\Delta}\) is the \(N \times N\) COSA dissimilarity matrix with elements \(D_{ij}[\mathbf{W}]\) and \(\mathbf{D}(\mathbf{Z})\) is the Euclidean distance matrix derived from the  \(N \times p\)  configuration matrix \(\mathbf{Z}\) that contains coordinates for the objects in a \(p-\)dimensional representation space. An example of an algorithm that minimizes such a metric least squares loss function is the so-called SMACOF algorithm. The original SMACOF (Scaling by Maximizing a Convex Function) algorithm is described in De Leeuw and Heiser (1982). Later, the meaning of the acronym was changed to Scaling by Majorizing a Complicated Function in Heiser (1995). 

The Classical Scaling approach, also known as Torgerson-Gower scaling (Young and Householder 1938; Torgerson 1952; Gower 1966), minimizes a loss function (called STRAIN in Meulman 1986) defined on scalar products (\(\mathbf{ZZ'}\)) and not on distances \(\mathbf{D}(\mathbf{Z})\), and is written as
\bea
\label{ToGo}
\mbox{STRAIN}(\mathbf{Z})= \left| \left| (-\frac{1}{2}\mathbf{J}\bs{\Delta}^{2}\mathbf{J}) - \mathbf{ZZ'} \right| \right|^2
\eea
\noindent where $\mathbf{J} = \mathbf{I} - N^{-1}\mathbf{11'}$, a centering operator that is applied to \textit {squared} dissimilarities in $\bs{\Delta}^{2}$. 
 
\begin{figure}[ht]
\begin{center}
 \includegraphics[width=7cm,keepaspectratio]{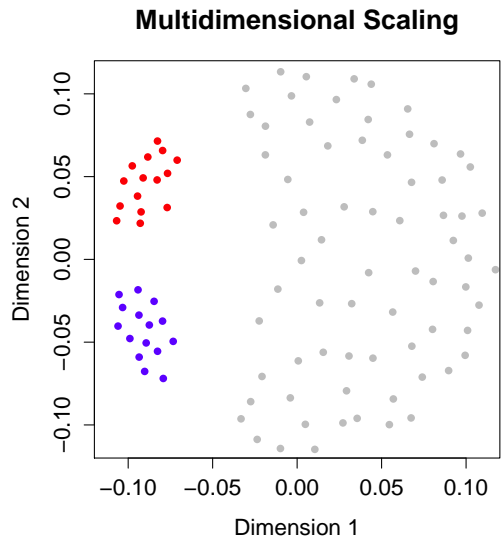}
\caption{Metric least squares multidimensional scaling solution}\label{fig:smacof_1}
\end{center}
\end{figure}

\begin{figure}[ht]
\begin{center}
\includegraphics[width=7cm,keepaspectratio]{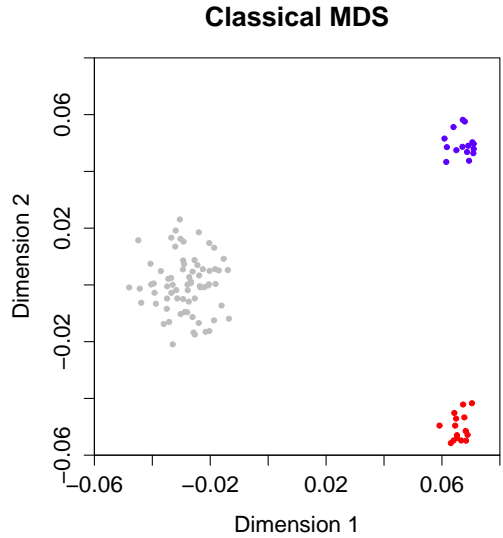} 
\caption{ Classical metric multdimensionaling solution}.
\label{fig:cmds_1}
\end{center}
\end{figure}

The drawback of minimizing the STRAIN loss function is that the resulting configuration $\mathbf{Z}$ is obtained by a projection of the objects into a low-dimensional space. Due to this projection, objects having distances that are large in the data, may be displayed close together in the representation space, giving a false impression of similarity. By contrast, a least squares metric MDS approach (such as SMACOF) gives a nonlinear mapping instead of a linear projection, and will usually preserve large distances in low-dimensional space. See Meulman (1986, 1992) for more details. 
In the following MDS applications, we will display objects in two-dimensional space, showing both the classical solution and the least squares solution, in Figure \ref{fig:smacof_1} and Figure \ref{fig:cmds_1}, respectively.

For the argument \texttt{groups} in the \texttt{smacof} function, we can use \texttt{grps.cosa} obtained from \texttt{getclust} to give different colors to points in the two groups.
\begin{Sinput}
> smacof.rslt <- smacof(cosa.rslt$D, groupnr = grps.cosa$grps, interc = 0)
\end{Sinput}
\FloatBarrier
\FloatBarrier
Figure \ref{fig:smacof_1} shows the metric least squares MDS solution for the two groups of objects (in red and blue), while the gray objects show a typical representation of a high-dimensional cloud of points with equal dissimilarities, nonlinearly mapped into two-dimensional space.

\begin{figure}[ht]
\begin{center}
\includegraphics[width=14cm,keepaspectratio]{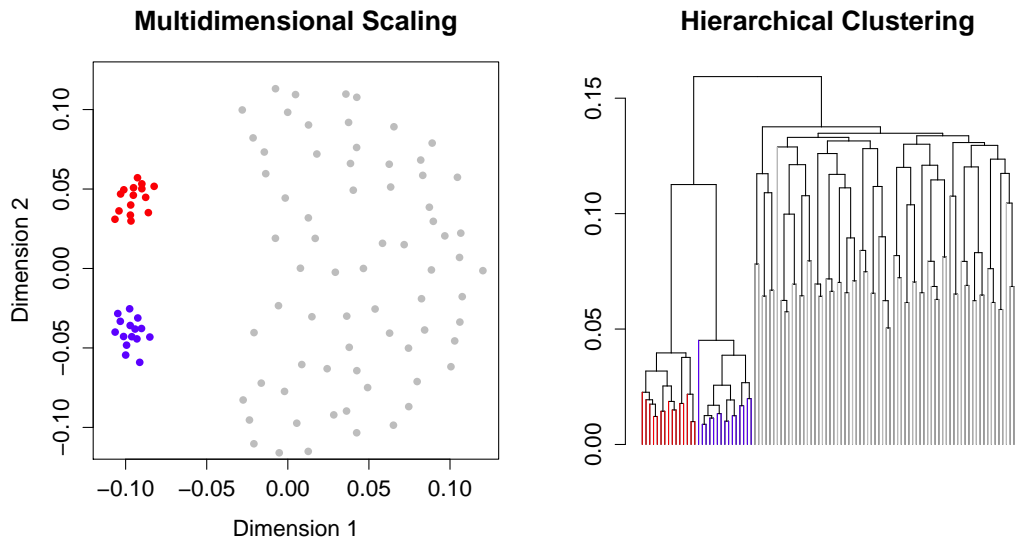} 
\caption{ Multidimensional scaling solution and dendrogram, after eliminating an additive constant}.
\label{fig:smacof_2}
\end{center}
\end{figure}

The \texttt{cmds} function in the \texttt{rCOSA} package, implementing the Classical Scaling procedure, is derived from the function \texttt{cmdscale} in the \texttt{stats} package. By using the commands
\begin{Sinput}
> xclas <- cmds(cosa.rslt$D, groupnr = grps.cosa$grps)
\end{Sinput}
we obtain Figure \ref{fig:cmds_1}. We observe that the large cloud of gray points, representing objects that are not similar to any of the other objects, seem to form a cluster as well; this is undesirable, since they are noise objects. Their closeness is due to the linear projection characteristic for classical MDS. Therefore, the representation given by the \texttt{smacof} function, given in Figure \ref{fig:smacof_2}, is to be preferred since it shows that the noise objects are not closely related.

\FloatBarrier
\FloatBarrier
At this point, we should address the possibility of having an additive constant present in the standard dissimilarity output of COSA in \texttt{cosa.rslt\$D}. We can take care of such a constant by fitting an \textit{interval} transformation to the COSA dissimilarities $D_{ij}[\mathbf{W}]$:
\[
{\hat{d}}_{ij}=\alpha + \beta D_{ij}[\mathbf{W}],
\]
taking care that ${\hat{d}}_{ij}$ does not become negative. We do this by using the \texttt{intercept} option (by default set to 1) in the \texttt{smacof} function:   
\begin{Sinput}
> smacof.rslt <- smacof(cosa.rslt$D, groupnr = grps.cosa$grps, interc = 1)
\end{Sinput}
When we iteratively minimize the so-called nonmetric least squares loss function over \(\hat{d}_{ij}\) and \(\mathbf{Z}\) 
\[ \mbox{STRESS} (\mathbf{Z}) = \sum_{i = 1}^N \sum_{j = 1}^N ( \hat{d}_{ij} - d_{ij}(\mathbf{Z}))^2,\]
we obtain the representation of object points in Figure \ref{fig:smacof_2}.
\FloatBarrier
\FloatBarrier
We observe that the clusters are much tighter after eliminating the additive constant. Figure \ref{fig:smacof_2} also displays the corresponding dendrogram; the different coloring in the dendrogram has been obtained by using the command
\begin{Sinput}
> hclst.cosa <- hierclust(smacof.rslt$D, groupnr = grps.cosa$grps)
\end{Sinput}

\subsection{Using COSA with targeting}
\begin{figure}[ht]
\begin{center}
\includegraphics[width=14cm,keepaspectratio]{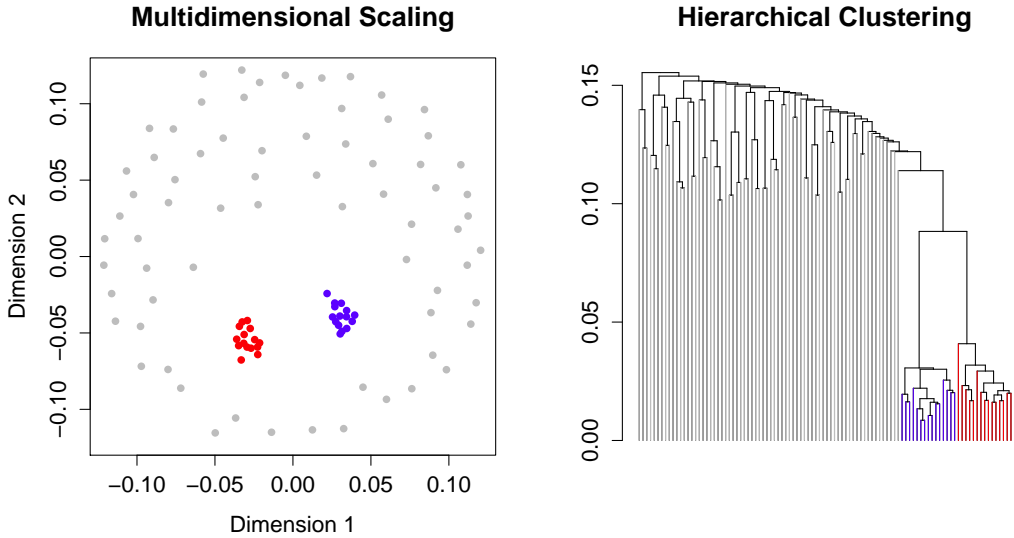} 
\caption { COSA with targeting: Multidimensional scaling solution and dendrogram, after eliminating an additive constant}.
\label{fig:smacof_3}
\end{center}
\end{figure}

To demonstrate the power of targeting in COSA, we analyze the same data set as in the previous section, but now using the commands
\begin{Sinput}
> cosa.rslt <- cosa2(X, targ = "high/low")
> smacof.rslt <- smacof(cosa.rslt$D, groupnr = grps.cosa$grps, interc = 1)
> hclst.cosa <- hierclust(smacof.rslt$D, groupnr = grps.cosa$grps)
\end{Sinput}
Since the design (see Figure \ref{fig:ExDesign2Intro}) created groups with both high (+1.5) and low(-1.5) values, we use double targeting ("high/low"). The results are given in Figure \ref{fig:smacof_3}.
\FloatBarrier
\FloatBarrier
Both representations of the COSA dissimilarities in Figure \ref{fig:smacof_3} clearly show that the distinction between the clusters on the one hand and the noise objects on the other hand have become much sharper.   
\subsection{Attribute Importance}

After having found clusters of objects in the data, we wish to know which attributes are important for the different clusters. The importance \(I_{kl}\) of attribute \(k\) for cluster \(l\) $(C_l)$ is inversely proportional to the dispersion \(S_{kl}\) of the data in attribute $k$ for objects in cluster $C_l$ of size $N_l$, and is defined as

\be
S_{kl} = \frac{1}{N^2}\sum_{i,j \in C_l} d_{ijk} \propto I_{kl}^{-1}.
\label{COSAimp}
\ee
\noindent If the dispersion of the data in an attribute is small for a particular group of objects, than the attribute is important for that particular group. Because the importance value is inversely proportional to within-group dispersion, the importance value is biased towards the variables with small within-group variability, and not towards large between-group separation. 

To see whether the value of a particular attribute importance is higher than could be expected by chance, a simple resampling method can be used. First, to determine how many attributes are important for a particular cluster, e.g., cluster $l$ of size $N_l$, we execute the commands
\begin{Sinput}
> attimp1.cosa <- attimp(X, group = grps.cosa$index$grp1, range = 1:1000)
> str(attimp1.cosa)
\end{Sinput}
The indices of the ordered attributes are given in \texttt{attimp1.cosa\$att}, and the corresponding descending attribute importance values in \texttt{attimp1.cosa\$imp}. To get a complete overview of the attribute importances, use the \texttt{attimp} function for the other groups as well.

\begin{Sinput}
> par(mfrow = c(3, 1))
> attimp1.cosa <- attimp(X, group = grps.cosa$index$grp1, range = 1:1000)
> attimp2.cosa <- attimp(X, group = grps.cosa$index$grp2, range = 1:1000)
> indx0 <- (1:N)[-c(grps.cosa$index$grp1, grps.cosa$index$grp2)]
> attimp0.cosa <- attimp(X, group = indx0, range=1:1000)
> par(mfrow = c(1,1))
\end{Sinput}

\begin{figure}[t]
\begin{center}
\includegraphics[]{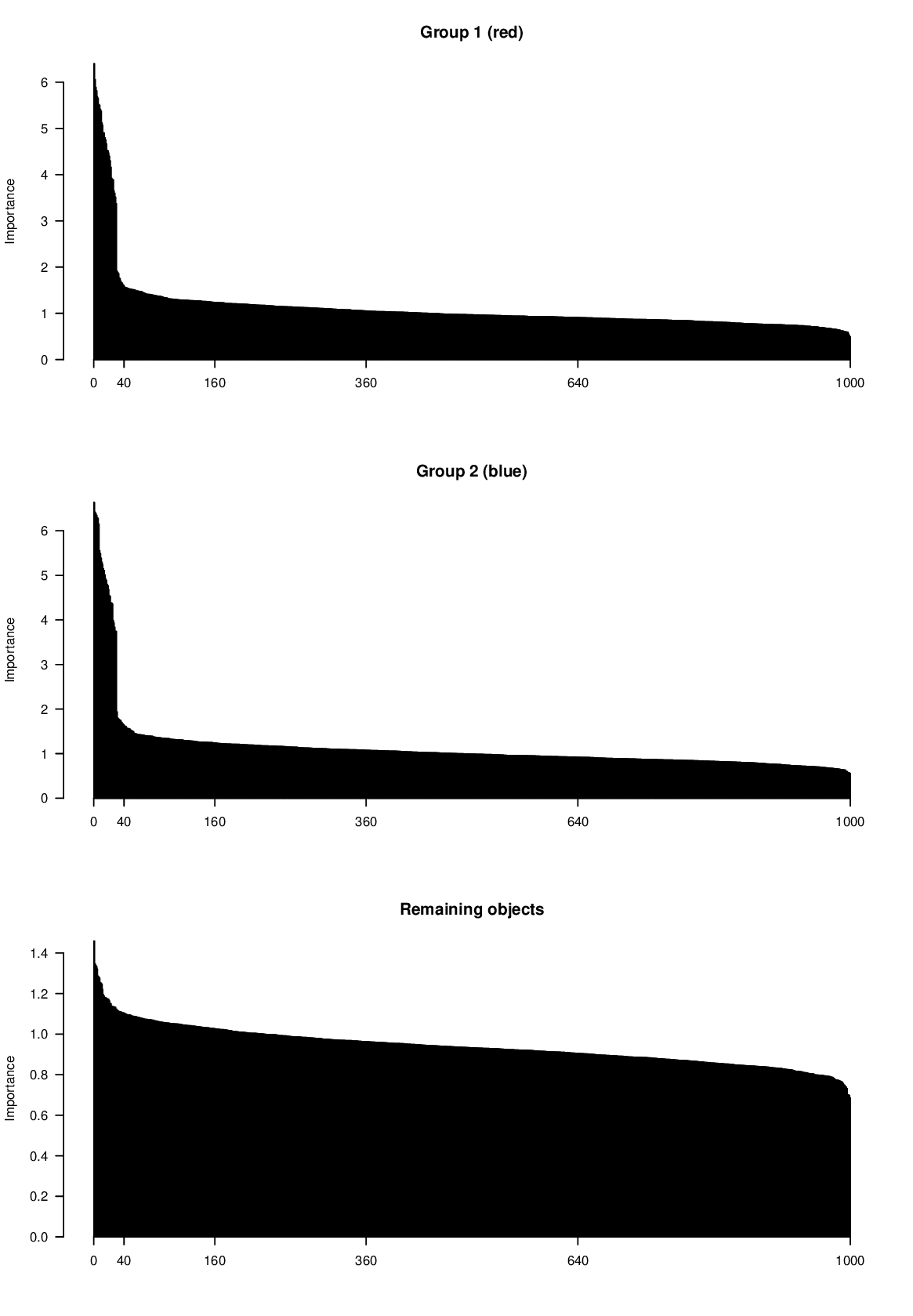}
\caption{ Display of the attribute importances of group 1, group 2 and the remaining objects in barplots.
\itc{}}\label{fig:importance}
\end{center}
\end{figure}

Note the differences in scale on the vertical axes for the groups in Figure \ref{fig:importance}. Having a good overview of the number of important attributes per group, we can obtain the maximum of the importance values and select the number of attributes that should be inspected according to their importance. Based on the above overview, we would select the first 50 attributes. Next, execute \texttt{attimp} again, now with 
\begin{Sinput}
> attimp1.cosa <- attimp(X, ylim = c(0,7), group = grps.cosa$index$grp1,
+   range = 1:50, times = 10, main = "Group 1 (Red)")
\end{Sinput}
By using these options, \texttt{attimp} will plot the 50 highest attribute importance values for cluster $l$, and will also take a random sample of size $N_l$ from the data for the first $50$ ordered attributes, and compute the attribute importance values on the basis of this random group. This is repeated 10 times. Also, note that we know the maximum of the importance values at this point, so we can set the limits of the vertical axes equal to each other for each group.
 
\begin{Sinput}
> lmts <- range(cosa.rslt$W[, k[1:50]]) # limits for the vertical axis
> par(mfrow = c(3,1))
> boxw(W = cosa.rslt$W, grpnr = grps.cosa$index$grp1, attr = k[1:50],
+     pch = ".", col = 'red', ylim = lmts, outline = F, xlab = "attributes", 
+     main = 'Group 1', ylab = 'weight value')
> boxw(W = cosa.rslt$W, grpnr = grps.cosa$index$grp2, attr = k[1:50],
+     pch = ".", col = 'red', ylim = lmts, outline = F, xlab = "attributes", 
+     main = 'Group 2', ylab = 'weight value')
> boxw(W = cosa.rslt$W, grpnr = indx0, attr = k[1:50],
+      pch = ".", col = 'gray', ylim = lmts, outline = F, xlab = "attributes", 
+      main = "Remaining objects", ylab = 'weight value')
> par(mfrow = c(1,1))
\end{Sinput}

\begin{figure}[t]
\begin{center}
\includegraphics[height=20cm]{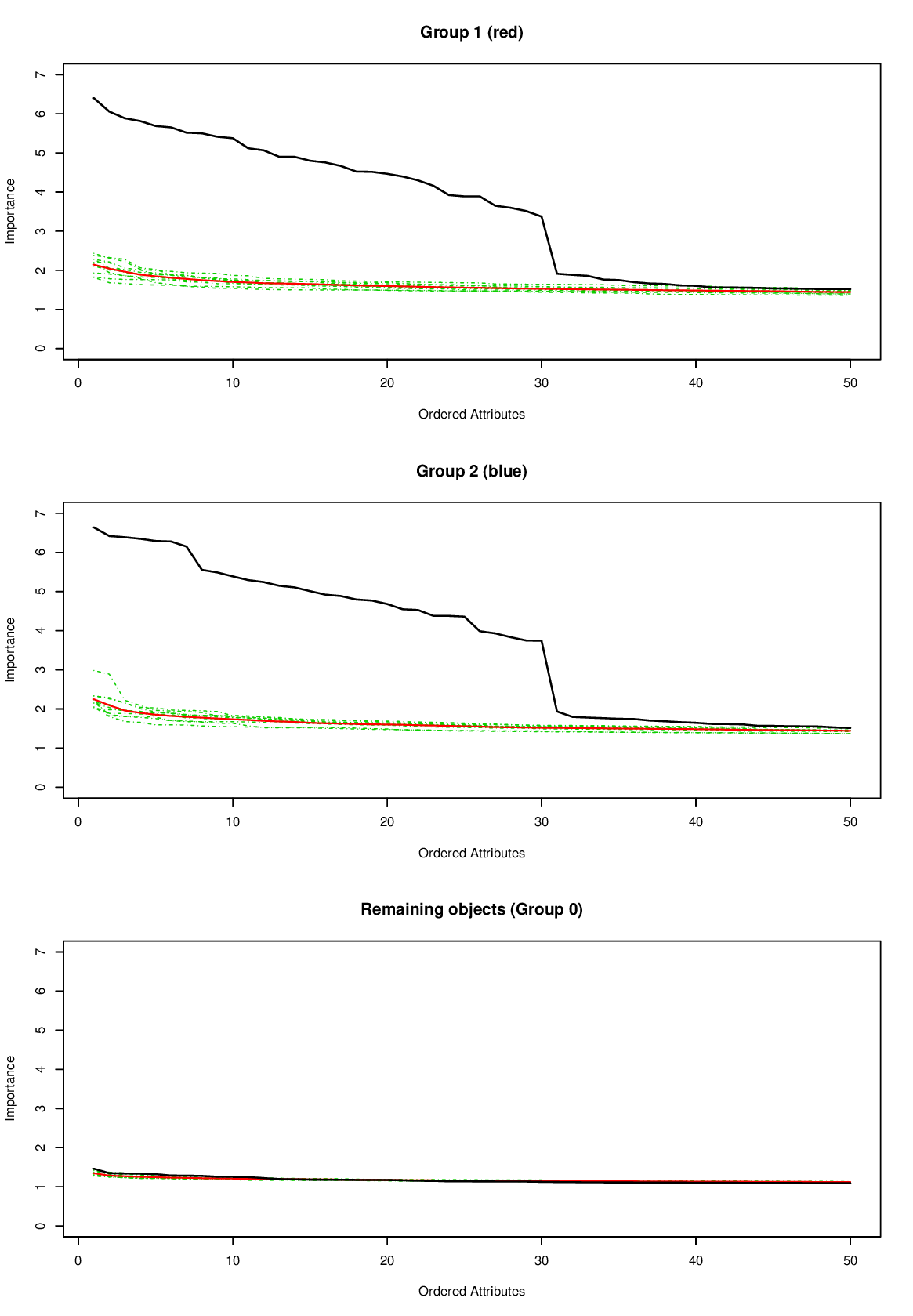}
\caption{Display of the attribute importances of group 1, group 2 and the remaining objects.
\itc{}}\label{fig:importance}
\end{center}
\end{figure}
\bigskip
In Figure \ref{fig:importance} the black line indicates the attribute importance of the attributes for each cluster. The green lines are the attribute importance lines for groups of the same size, randomly sampled from the data. The red line is the average of the green lines. Thus, the larger the difference between the black line and the red line, the more evidence that the attribute importance values are not just based on chance. Note the sudden drop of the black attribute importance line after 30 attributes. This is in line with the simulated data, in which each group is clustered on 30 attributes only. There are no attributes that can be considered important for the remaining objects.

In addition to the attribute importance values, we can also look at the attribute weight matrix, \texttt{cosa.rslt\$W}, and plot the values of the weights. For each group we use the first k'=1:50 weights to draw boxplots. The first 45 of these 50 weights are the important weights according to the design from Figure \ref{fig:ExDesign2Intro}. The code to draw the boxplots:

\begin{Sinput}
> par(mfrow = c(3,1))
> boxw(W = cosa.rslt$W, grpnr = grps.cosa$index$grp1, attr = k[1:50], 
+ 	pch = ".", col = 'red', ylim = lmts, outline = F)
> boxw(W = cosa.rslt$W, grpnr = grps.cosa$index$grp2, attr = k[1:50], 
+ 	pch= ".", col = 'blue', ylim = lmts, outline = F)
> boxw(W = cosa.rslt$W, grpnr = indx0, attr = k[1:50], 
+ 	pch = ".", col = 'gray',ylim = lmts, outline = F)
> par(mfrow = c(1,1))
\end{Sinput}
\begin{figure}[t]
\begin{center}
\includegraphics[height=20cm]{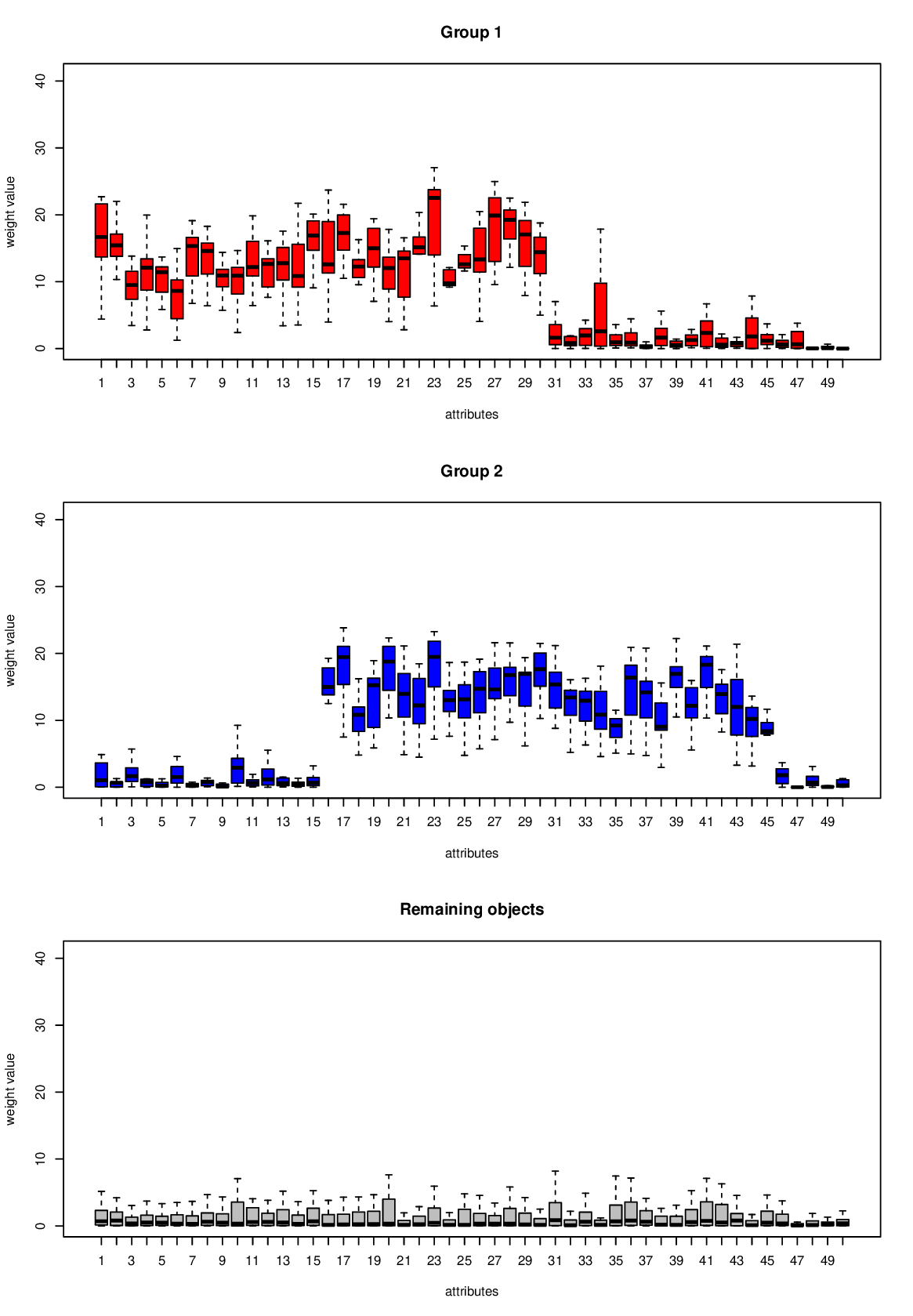}
\caption{ Boxplots of the weights of attributes $k' = \{1,\ldots 50\}$ for group 1, group 2 and the remaining objects.
\itc{}}\label{fig:BoxplotWeights}
\end{center}
\end{figure}
%
It is clear that the COSA weights display the same structure as was found for the attribute importances: Group 1 has large weights for attributes 1:30, group 2 has large weights for attributes 15:45, group 1 and 2 have large weights on the overlapping attributes 15:30, and all weights for the remaining objects are small. COSA clearly separates the signal from the noise in our data.

\begin{figure}[ht]
\begin{center}
\includegraphics[]{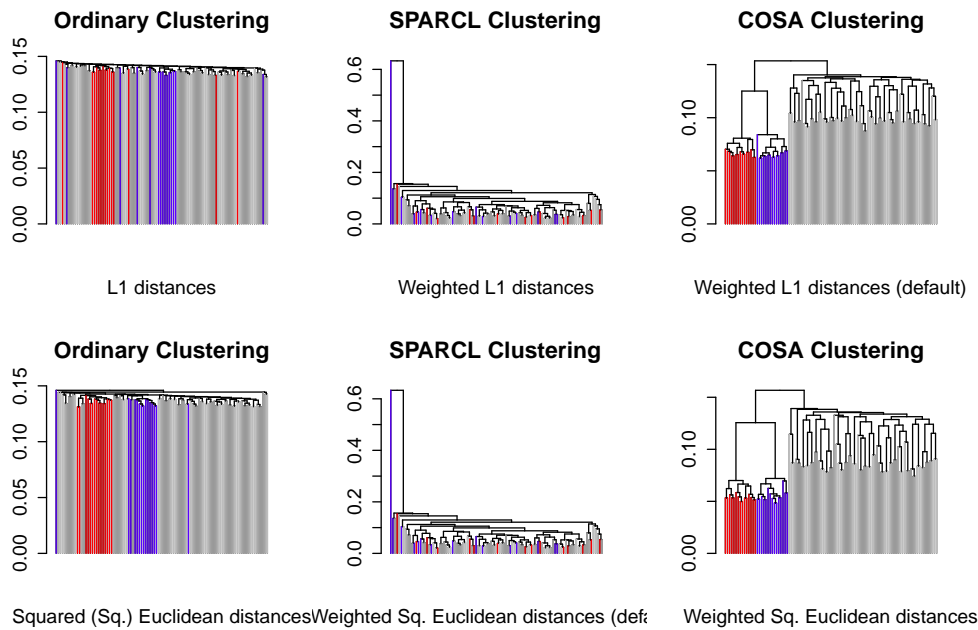}
\caption{Dendrograms obtained from hierarchical clustering for six different dissimilarity matrices derived from the
simulated data in Figure 2. $L_1$ distances in the first row, squared Euclidean distances in the second row. Unweighted dissimilarities in the first column, SPARCL dissimilarities in the second column, COSA dissimilarities in the third column.}
\label{fig:DendroMCstudy}
\end{center}
\end{figure}

Although the structure in the data was especially designed to demonstrate COSA, it is not particularly complicated. However, very common approaches in cluster analysis, such as hierarchical clustering of either squared Euclidean distances or $L_1$ distances, are not able to cope with it. This is also true for the more sophisticated SPARCL approach. Results are shown in Figure \ref{fig:DendroMCstudy}, where we give the dendrograms obtained for the COSA dissimilarities, the $L_1$ distances, the squared Euclidean distances, and the SPARCL dissimilarities, as defined in Equations \eqref{COSAd}, \eqref{DL1}, \eqref{DL2}, and  \eqref{SPARCLd}, respectively. To obtain the COSA and the SPARCL dissimilarities, we used the default settings, which amounts to weighted $L_1$ distances in COSA and weighted squared Euclidean dissimilarities in SPARCL.
\FloatBarrier
\FloatBarrier

\section{Analysis of the Leiden ApoE3 data}

The data in the following example are from an experiment with two types of mice: normal mice (called `wildtype') and transgenic mice. The latter type contains the Human Leiden ApoE3 variety. The biological background is briefly summarized as follows. ApoE3 stands for Apolipoprotein E3; it is one of many apolipoproteins that, together with lipids, form lipoproteins (cholesterol particles), for example, LDL, VLDL, and HDL. The E3 ``Leiden" is a human variant of ApoE3. When the lipoprotein is no longer recognized by special receptors in the liver, it prevents uptake of LDL cholesterol by the liver, and this results in strongly increased lipoprotein levels in the plasma. Eventually the latter condition results in atherosclerosis, which is hardening of the arteries. This may lead to blocked blood vessels and a stroke or a heart attack.
The experiment has two important features. Mice would usually develop severe atherosclerosis when on a high fat diet. However, in the current experiment, the mice were on a low fat diet. Also atherosclerosis would be manifest after 20 weeks, but the samples were collected when the mice were only 9 weeks of age.

\subsection{Data Description}
The 1550 attributes in the study are LC-MS (liquid chromatography-mass spectrometry) measurements of plasma lipids. The objects consist of 38 cases, with two observations for each mouse. The original experiment was performed with 10 wildtype and 10 transgenic mice, but only 9 transgenic mice survived the experiment (Damian, Oresics, Verheij, Meulman, Friedman, Adourian, Morel, Smilde and van der Greef 2007).

\subsection{COSA analysis}

\begin{figure}[ht]
\begin{center}
\includegraphics[]{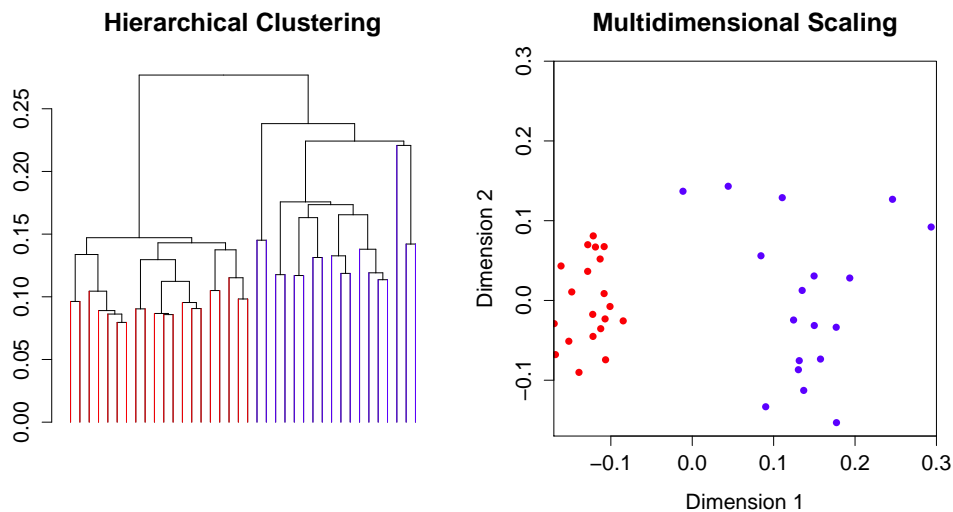}
\caption{Hierarchical cluster analysis and \texttt{smacof} of the COSA dissimilarities of the ApoE3 Leiden data. Wildtype mice are in red, and transgenic mice are in blue. Average linkage was used in the clustering.
\itc{}}\label{fig:DendroSmacofMICE}
\end{center}
\end{figure}

The COSA analysis consists of first computing the dissimilarity matrix based on the COSA weights, and then subjecting this matrix to hierarchical clustering 
(using \texttt{hierclust}) and multidimensional scaling (using \texttt{smacof}), resulting in a dendrogram and a two-dimensional space, respectively (shown in Figure \ref{fig:DendroSmacofMICE}).

\begin{Sinput}
> data(ApoE3)# load ApoE3 data from rCOSA package
>  par(mfrow = c(1,2))
>  cosa.AE3 <- cosa2(ApoE3)
>  hc.AE3 <- hierclust(cosa.AE3$D)
>  grps.AE3 <- getclust(hc.AE3) # select clusters
>  smacof_AE3 <- smacof(cosa.AE3$D, groupnr = grps.AE3$grps, 
+  	niter = 100, interc = 1)
\end{Sinput}

\FloatBarrier
\FloatBarrier

We have used the average link option (the default) for the hierarchical cluster analysis, but this choice was not essential for the separation between the transgenic and the wild type mice, which is perfect. 

Again, we use the \texttt{attimp()} function to inspect the importance values for the variables in each of the two clusters.

\begin{Sinput}
> par(mfrow = c(2,1))
> attWild <- attimp(ApoE3, group = grps.AE3$index$grp1, times = 10, 
+    main = "Wildtype Mice", range = 1:250, 
+    xlab = 'ordered attributes (the first 80 out of 1550)')
> attTrans <- attimp(ApoE3, group = grps.AE3$index$grp2, times = 10, 
+    main = "Transgenic Mice", range = 1:250, 
+    xlab = 'ordered attributes (the first 80 out of 1550)')
\end{Sinput}

In the ApoE3 data, only a small number (40-60) of the original 1550 attributes turn out the be important. Ten random groups of the size of group of the wild type cluster (which is 20) are sampled from the data, and for each of these random samples the importance values are computed. Then the actual importance values found are compared to those from the test, and in this way we can determine which variables are more important than can be attributed to chance. We also perform this test for the transgenic cluster. 

\FloatBarrier
\begin{figure}[ht]
\begin{center}
\includegraphics[width=14cm, keepaspectratio]{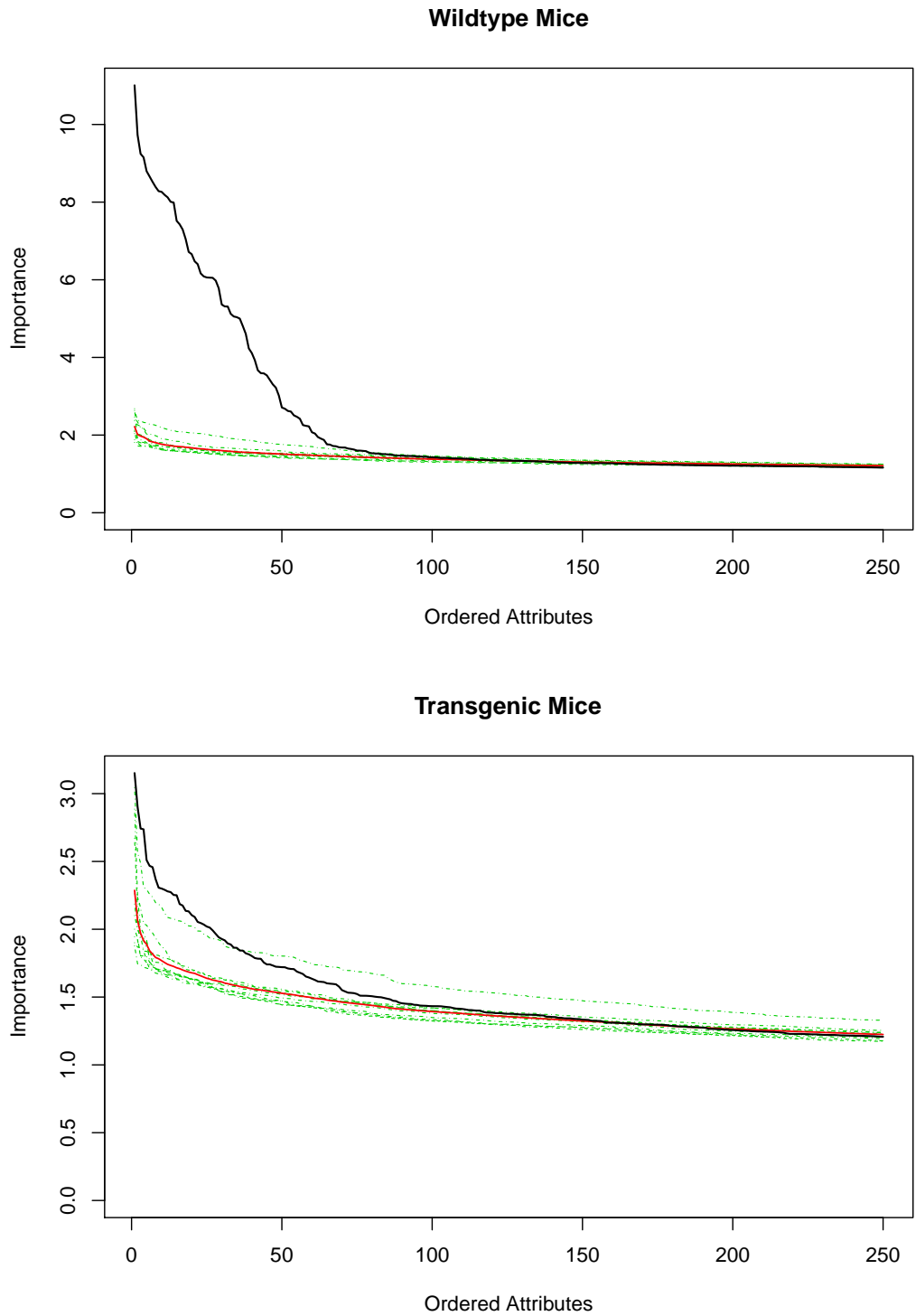}
\caption{ The two black curves in the upper and lower graph display the 85 largest (out of 1550) importance values for the group of transgenic mice 1-18 in the ApoE3 Leiden data (at the top) and for the wildtype mice (at the bottom). In each graph the ten green curves indicate the 85 largest importance values for ten random groups of size 18 and 20, respectively. The two red curves are the averages of each set of ten green curves.
\itc{}}\label{fig:AttimpMICE}
\end{center}
\end{figure}
\FloatBarrier

\begin{figure}[ht]
\begin{center}
\includegraphics[width=14cm]{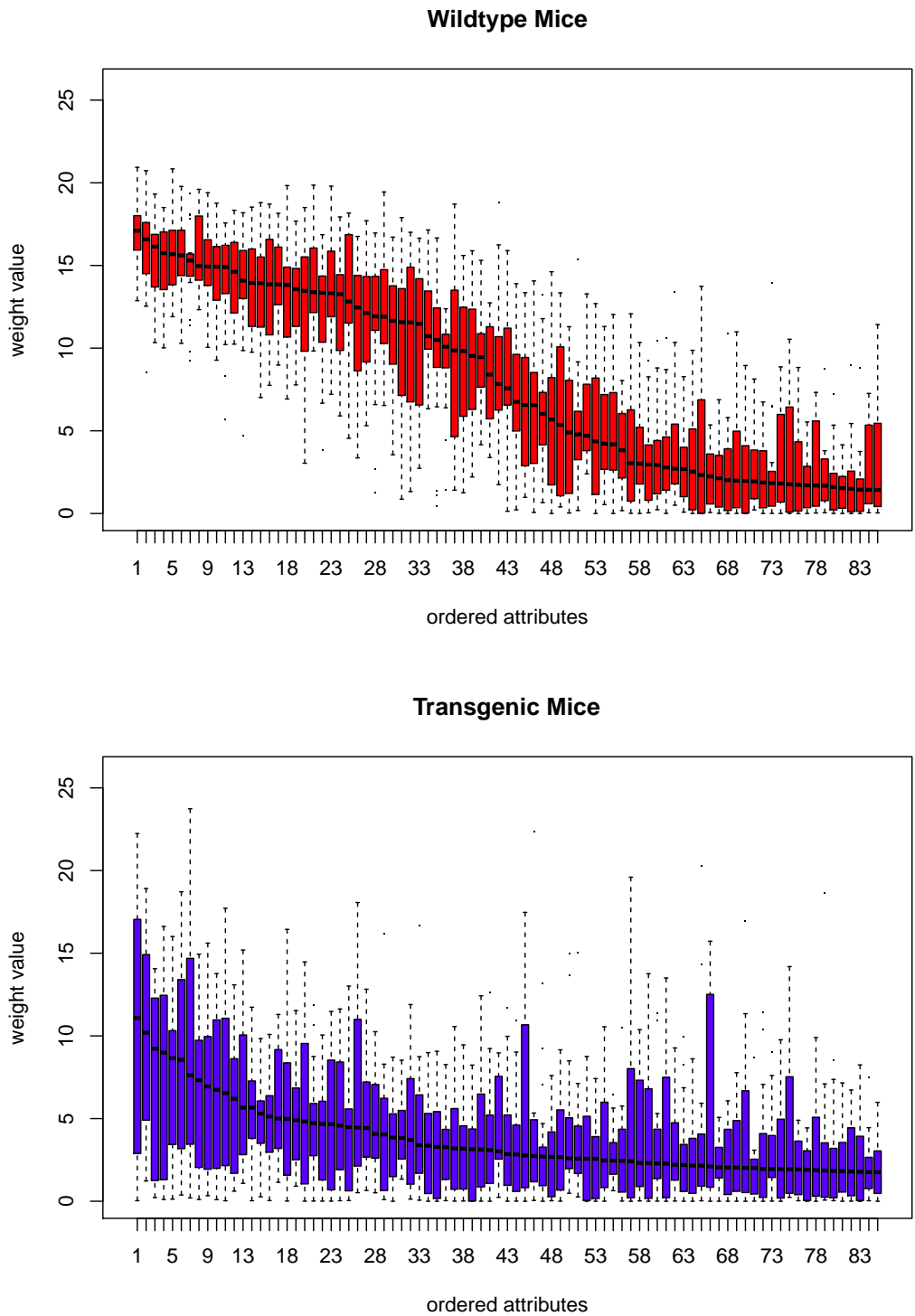}
\caption{ Boxplots for the weights of the median ordered attributes for the wildtype mice (top) and the weights of the median ordered attributes for the transgenic mice (bottom) separately.
\itc{}}\label{fig:BoxplotWeightsMICE}
\end{center}
\end{figure}

Here the values for the 85 most important variables (out of 1550) are displayed. The black curve gives the observed importance values, the ten green curves are for the randomly generated samples, and the red curve is again the average of the ten green curves. The difference between the importance values for the wildtype cluster and those for the 10 random groups is large; about 60 attributes appear to be important for the clustering of the wildtype group. The importance values for the transgenic cluster are somewhat less distinct. It is clear, however, that note more than 100 variables are truly important for the clustering of the transgenic mice.
We obtain boxplots (Figure \ref{fig:BoxplotWeightsMICE}) for the weights of the first 85 attributes, ordered from most to least important within each group, by:
\begin{Sinput}
> lmts <- range(cosa.AE3$W) # set the limits of the y-axis
> par(mfrow = c(2,1))
> boxw(W = cosa.AE3$W, 
+      grpnr = 19:38, 
+      attr = attWild$att[1:85], 
+      pch = ".", col = 'blue', ylim = lmts,
+      main = 'Wildtype Mice' )
> boxw(W = cosa.AE3$W, 
+      grpnr = 1:18, 
+      attr = attTrans$att[1:85], 
+      pch = ".", col = 'red', ylim = lmts,
+      main = 'Transgenic Mice' )
> par(mfrow = c(1,1))
\end{Sinput}

\FloatBarrier
\FloatBarrier
When we take a look at the boxplots of the attribute weights within each group, we can conclude that the medians of the weights of the ordered attributes in the wild type group are much more distinct compared to the the transgenic group.

\begin{figure}[ht]
\begin{center}
\includegraphics[]{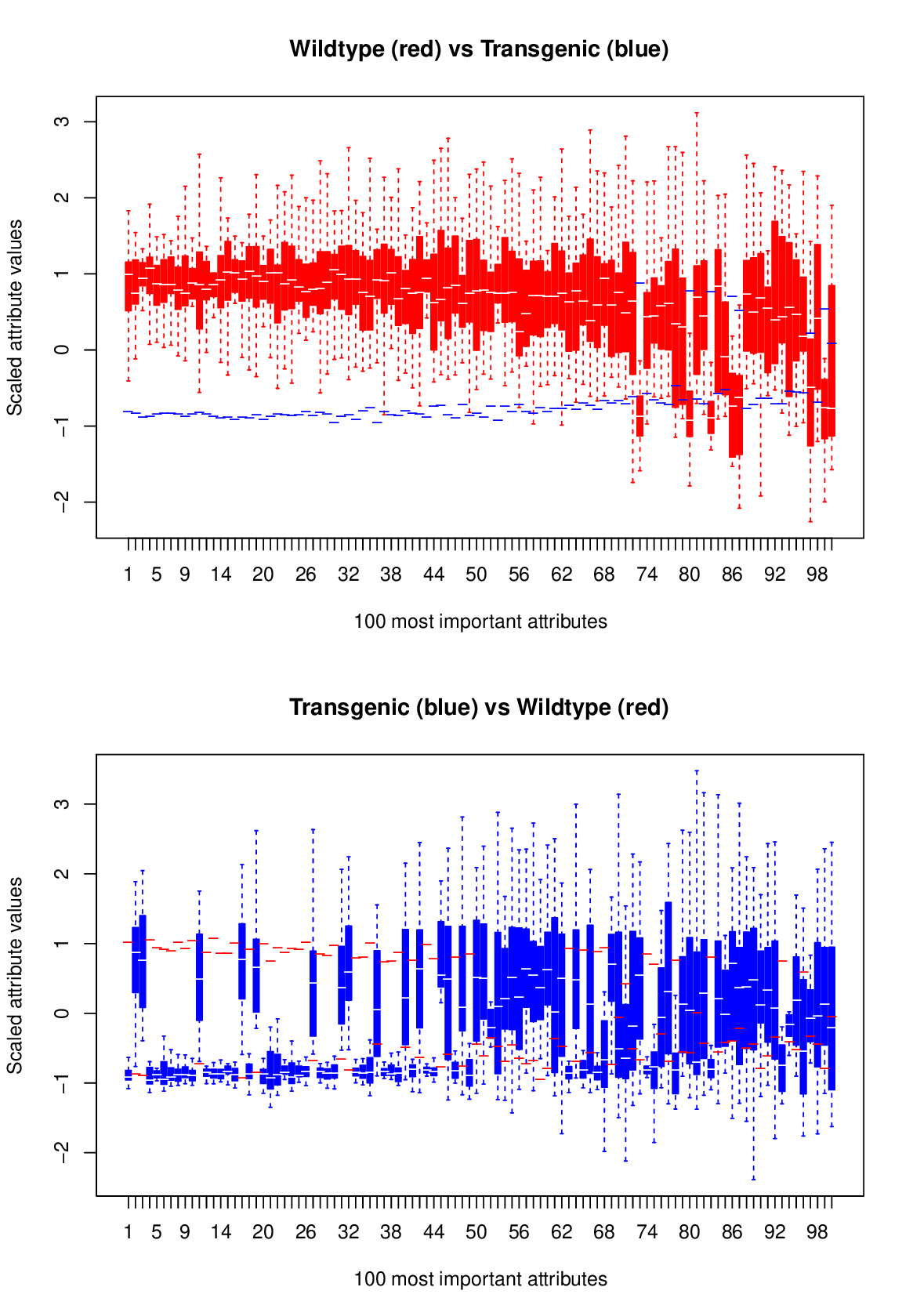}
\caption{In the top panel the values of the 100 most important attributes for the wildtype group are shown in red boxplots. In blue the median values of the transgenic group are added for these attributes. In the bottom panel the values of the 100 most important attributes for the transgenic group are shown in blue boxplots. In red the median values of the wildtype group are added for these attributes.}\label{fig:ApoE3Values}
\end{center}
\end{figure}

\begin{figure}[ht]
\begin{center}
\includegraphics[]{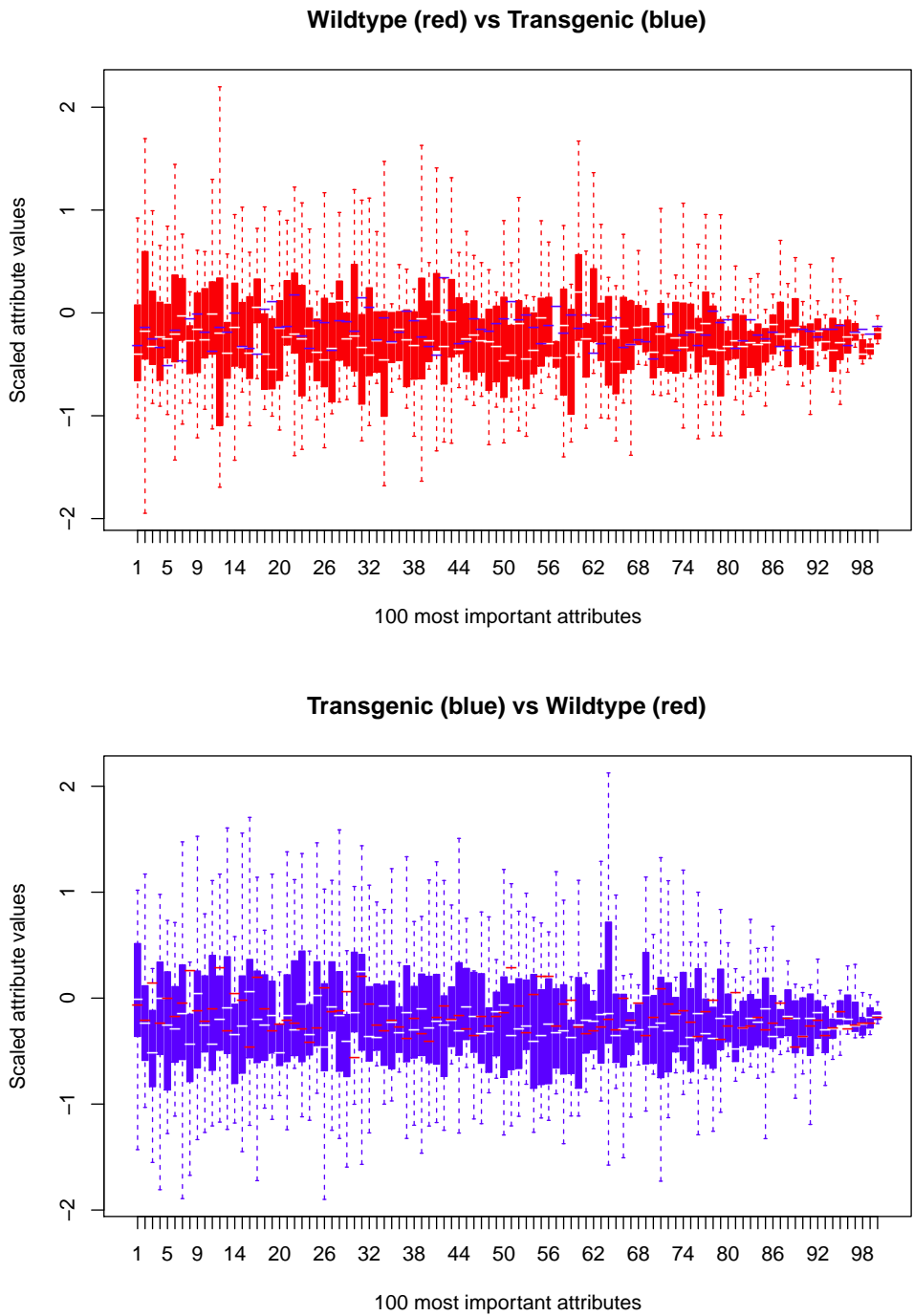}
\caption{In the top panel the values of the 100 least important attributes for the wildtype group are shown in red boxplots. In blue the median values of the transgenic group are added for these attributes. In the bottom panel the values of the 100 least important attributes for the transgenic group are shown in blue boxplots. In red the median values of the wildtype group are added for these attributes.}\label{fig:ApoE3Values2}
\end{center}
\end{figure}

In Figures \ref{fig:ApoE3Values} and \ref{fig:ApoE3Values2}, we inspect the attribute values for the 100 most and least important attributes. Values for the wildtype group are ordered according to attribute importance, and are contrasted with their corresponding attribute median values for the transgenic group and vice versa.
\FloatBarrier
\FloatBarrier

The \texttt{R} code to generate the plots in Figure \ref{fig:ApoE3Values} is:
\begin{Sinput}
par(mfrow = c(2,1))  
> boxatt(data = ApoE3, imps = attWild$att[1:100], 
+      grps =  list(1:18, 19:38), 
+       main = "Wildtype (red) vs Transgenic (blue)", 
+       ylab = 'Scaled attribute values', 
+      xlab = '100 most important attributes',
+      colors = c('red','white','blue'))
> boxatt(data = ApoE3, imps = attTrans$att[1:100], 
+       grps =  list(19:38,1:18), 
+       main = "Transgenic (blue) vs Wildtype (red)", 
+       ylab = 'Scaled attribute values', 
+       xlab = '100 most important attributes',
+       colors = c('blue','white','red'))
> par(mfrow = c(1,1))
\end{Sinput}

and for Figure \ref{fig:ApoE3Values2}, we simply replace the indices \texttt{1:100} by \texttt{1451:1550}. 

\begin{figure}[ht]
\begin{center}
\includegraphics[]{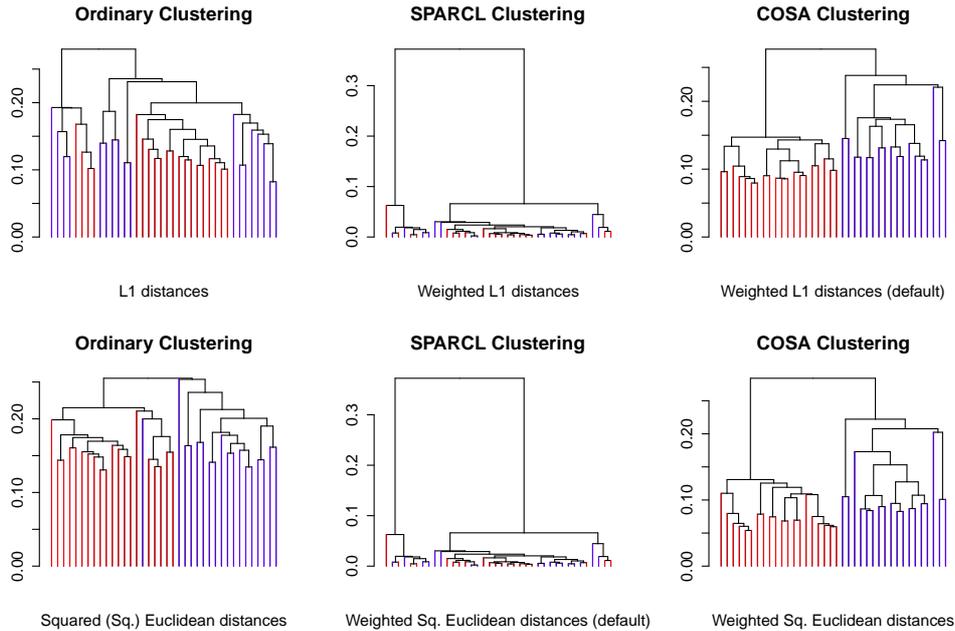}
\caption{Dendrograms obtained from hierarchical clustering for six different dissimilarity matrices derived from the ApoE3 mice data. $L_1$ distances in the first row, squared Euclidean distances in the second row. Unweighted dissimilarities in the first column, SPARCL dissimilarities in the second column, COSA dissimilarities in the third column.}\label{fig:ApoE3DendroComp}
\end{center}
\end{figure}

Finally, Figure \ref{fig:ApoE3DendroComp} displays six dendrograms obtained from hierarchical clustering of dissimilarities derived from the ApoE3 Mice data.  As in Figure \ref{fig:DendroMCstudy} for the simulated data, we have used both ordinary $L_1$ and squared Euclidean distances, and weighted SPARCL and COSA distances. The object structure in the ApoE3 Mice data is very different from the simulated data, because the latter contain two small groups of 15 objects and 70 noise objects.  The ApoE3 Mice data do not contain noise objects; the 38 objects come from two experimental conditions, with two groups of 20 and 18 objects, respectively. As we see in Figure \ref{fig:ApoE3DendroComp} , the dendrograms for ordinary clustering based on $L_1$ and squared Euclidean distances (left panels) are far from perfect, although the first is better than the second. Based on inspection of the dendrograms, COSA with squared Euclidean distances performs somewhat better than COSA with $L_1$ distances (right panels). Surprisingly, neither variant of SPARCL does perform well for these data. Even more unexpected, is the performance of SPARCL compared to the ordinary clustering dendrograms in the left panels, where all attributes equally contribute to the dissimilarities.

\FloatBarrier
\FloatBarrier

%
%
%

\section{Discussion}

We demonstrated with two examples the use of the new software package \texttt{rCOSA}; the first was a simulated data set, and the second a complex metabolomics data set. Compared to other commonly used distance methods, COSA was shown to be very powerful in retrieving and revealing a cluster structure. When using the current default settings, only very few \texttt{R} command line skills are needed to use the \texttt{rCOSA} package. 

Those with some extended \texttt{R} programming skills, can use the output from the \texttt{cosa2} for further use. Analysis of the COSA dissimilarity matrix is not limited to hierarchical clustering or multidimensional scaling, as was presented in this tutorial. Other linear or non-linear projection methods that use dissimilarity matrices, such as self organizing maps (Kohonen 2001), Sammon's mapping (Sammon 1969) or curvilinear distance analysis (Lee, Lendasse and Verleysen 2004), may also be considered. Compositional data analysis (Aitchison 1986) of the COSA weight matrix may also lead to additional insights in the cluster structure of the objects.


To date, we are not aware of any other software interfaced to \texttt{R} that outputs dissimilarities for clustered objects on (different) subsets of attributes. Packages that we found all assume the number of clusters to be set beforehand. Examples of these are \texttt{ORCLUS} (Szepannek 2013), \texttt{wksm} (Williams, Huang, Chen, Wang and Xiao 2014), and \texttt{FisherEM} (Bouveyron and Brunet 2012). COSA's strength is foremost derived from its capacity to find an unknown number of clusters, possibly among a large number of unclustered objects, where each cluster is associated with its own subset of important attributes.

%

\bibliographystyle{ECA_jasa}

\def\bibindent{1em}

% ...

\end{document}